\documentclass{article}

\usepackage{arxiv}

\usepackage[utf8]{inputenc} 
\usepackage[T1]{fontenc}    
\usepackage{hyperref}       
\usepackage{url}            
\usepackage{booktabs}       
\usepackage{amsfonts}       
\usepackage{nicefrac}       
\usepackage{microtype}      
\usepackage{lipsum}
\usepackage{graphicx}
\usepackage{natbib}
\usepackage{float}
\RequirePackage[T1]{fontenc}
\RequirePackage[utf8]{inputenc}
\RequirePackage{calc}
\RequirePackage{indentfirst}
\RequirePackage{fancyhdr}
\RequirePackage{graphicx,epstopdf}
\RequirePackage{lastpage}
\RequirePackage{ifthen}
\RequirePackage{float}
\RequirePackage{amsmath}
\RequirePackage{amssymb} 
\RequirePackage[right]{lineno}
\RequirePackage{setspace}
\RequirePackage{enumitem}
\RequirePackage{mathpazo}
\RequirePackage{booktabs} 
\RequirePackage{titlesec}
\RequirePackage{etoolbox} 
\RequirePackage{tabto} 
\RequirePackage{xcolor, colortbl} 
\RequirePackage{soul} 

\RequirePackage{multirow}
\RequirePackage{microtype} 
\RequirePackage{tikz} 
\RequirePackage{refcount} 
\RequirePackage{changepage} 
\RequirePackage{attrib} 
\RequirePackage{upgreek} 
\RequirePackage{array} 
\RequirePackage{tabularx}
\RequirePackage{pbox} 
\RequirePackage{ragged2e} 
\RequirePackage[subfigure]{tocloft} 
\RequirePackage{marginnote} 
\reversemarginpar 
\RequirePackage{marginfix} 
\RequirePackage{enotez} 
\RequirePackage{xstring} 
\RequirePackage[labelformat=simple]{subfig} 
\graphicspath{ {./images/} }
\usepackage[utf8]{inputenc}  
\usepackage[T1]{fontenc}
\usepackage{lmodern}
\usepackage{authblk}   

\title{Rule-based Generation of de Bruijn Sequences: Memory and Learning}

\author[1]{Francisco J. Mu\~noz}
\author[2]{Juan Carlos Nu\~no}

\affil[1]{Departamento de Matem\'atica Aplicada, Ciencia e Ingenier\'ia de Materiales y Tecnolog\'ia Electr\'onica\\
Universidad Rey Juan Carlos, 28933 Madrid, Spain\\
\texttt{francisco.munoz@urjc.es}}

\affil[2]{Departamento de Matem\'atica Aplicada\\
Universidad Polit\'ecnica de Madrid, 28040 Madrid, Spain\\
\texttt{juancarlos.nuno@upm.es}}

\begin{document}
\maketitle
\begin{abstract}
We investigate binary sequences generated by non-Markovian rules with memory length $\mu$, similar to those adopted in elementary cellular automata. This generation procedure is equivalent to a shift register, and certain rules produce sequences with maximal periods, known as de Bruijn sequences. We introduce a novel methodology for generating de Bruijn sequences that combines (i) a set of derived properties that significantly reduce the space of feasible generating rules and (ii) a neural-network-based classifier that identifies which rules produce de Bruijn sequences. The experiments for some values of $\mu$ demonstrate the approach’s effectiveness and computational efficiency.
\end{abstract}


\section{Introduction}

Cellular automata (CA) are a class of dynamical systems that evolve in discrete time and space \citep{Wolfram1983,Wolfram1984}. A CA consists of cells (or agents), each of which can adopt a state from a discrete set. The most commonly used alphabet is binary, typically denoted by $\{0,1\}$. The CA evolves by applying generating rules at each time step. These rules can be static or dynamic (i.e., they may change depending on the current state of the entire CA) and either local or global, depending on whether they involve only neighboring cells or the \mbox{entire population.}

From an initial configuration, the state of the CA evolves over time according to the specified rules. The update rule for each cell can depend on both the past states of the cell itself (temporal dimension) and the states of its neighboring cells (spatial dimension). Elementary one-dimensional cellular automata (1dCA) typically assume a memoryless structure with a linear topology, where each cell interacts only with its nearest neighbors \citep{Wolfram1983}. More complex models incorporate memory—i.e., dependence on a set of past states—and long-range spatial influences \citep{Li, Ramon}.

Relatively less attention has been given to zero-dimensional cellular automata (0dCA), which consist of a single cell whose state is updated based solely on its own past states. Despite their apparent simplicity, 0dCA do not involve interactions between multiple cells,  their time evolution yields interesting sequences of symbolic states. The rules governing such systems depend on the memory $\mu$—the number of previous states influencing the next state—and the alphabet $\{\alpha\}$, the number of possible states a cell can take. Generally, the complexity of the system increases with both $\mu$ and $\alpha$. For simplicity, we focus on binary systems, i.e., $\alpha=2$, using the symbols $\{0,1\}$, and, due to computational limitations, we restrict our study to a subset of cases with relatively low values of $\mu$ (less than 8).

\section{0-Dimensional CA}\label{0d}

Symbolic sequences naturally arise in discrete dynamical systems, e.g., CA \citep{Jin}. From an initial condition, the next state is computed according to a generation rule (or function) that depends on previous states. For instance, a classical non-Markovian binary time series can be defined by initial values \( a_1 \) and \( a_2 \) and a recursive {rule:} 
\begin{equation}
a_{k+2} = (a_{k+1} + a_k) \mod 2, \quad \text{for } k = 1,2,\ldots
\end{equation}

With initial states \( a_1 = 0 \) and \( a_2 = 1 \), the sequence continues as \( a_3 = 1 \), \( a_4 = 0 \), \( a_5 = 1 \), \( a_6 = 1 \), etc., producing the following binary sequence:
\[
s_1 = 0\, 1\, 1\, 0\, 1\, 1\, 0\, 1\, 1\, 0\, 1\, 1\, 0\, \ldots
\]
This sequence is clearly periodic with period \( T_1 = 3 \). In this case, the next digit depends on the two previous digits, implying a memory \( \mu = 2 \).

Other examples with \( \mu = 2 \) can be constructed using logical operators. For instance, applying the AND-function with the rules  
\[
1\,1 \rightarrow 1, \quad 1\,0 \rightarrow 0, \quad 0\,1 \rightarrow 0, \quad 0\,0 \rightarrow 0
\]
to the initial pair \( 0\,1 \) produces the sequence
\[
s_2 = 0\, 1\, 0\, 0\, 0\, 0\, 0\, \ldots
\]
Similarly, applying the OR-function
\[
1\,1 \rightarrow 1, \quad 1\,0 \rightarrow 1, \quad 0\,1 \rightarrow 1, \quad 0\,0 \rightarrow 0
\]
to the same initial configuration \( 0\,1 \) results in
\[
s_3 = 0\, 1\, 1\, 1\, 1\, 1\, 1\, \ldots
\]
Both sequences converge to fixed values (0 or 1), i.e., their periods are \( T_2 = T_3 = 1 \), and these fixed points are independent of the initial configuration.

In contrast, applying the XOR-function
\[
1\,1 \rightarrow 0, \quad 1\,0 \rightarrow 1, \quad 0\,1 \rightarrow 1, \quad 0\,0 \rightarrow 0
\]
to the initial state \( a_1 = 0, a_2 = 1 \) yields
\[
s_4 = 0\, 1\, 1\, 0\, 1\, 1\, 0\, 1\, 1\, 0\, \ldots
\]
with period \( T_4 = 3 \). However, using the same rule with the initial configuration \( a_1 = 0,\) \( a_2 = 0 \) results in a constant sequence \( a_k = 0 \) for all \( k > 2 \). These examples illustrate that the resulting behavior depends not only on the rule but also on the initial conditions.

For any finite memory \( \mu < \infty \), all sequences generated in this framework are eventually periodic. That is, for any rule and initial condition, the sequence will eventually settle into a repeating pattern (also referred to as a motif or loop). The maximum possible period for a sequence with memory \(\mu\) is
\begin{equation}
T_{\text{max}} = 2^\mu
\end{equation}

As \(\mu\) increases, the sequences may resemble aperiodic ones, making them useful for computing pseudorandom numbers to be applied in Monte Carlo simulations and cryptographic systems \citep{Etzion}. Truly aperiodic behavior can only arise when \(\mu \rightarrow \infty\), or when the alphabet is infinite (e.g., using real-valued states).

\begin{figure}[H]
\includegraphics[width=0.85\textwidth]{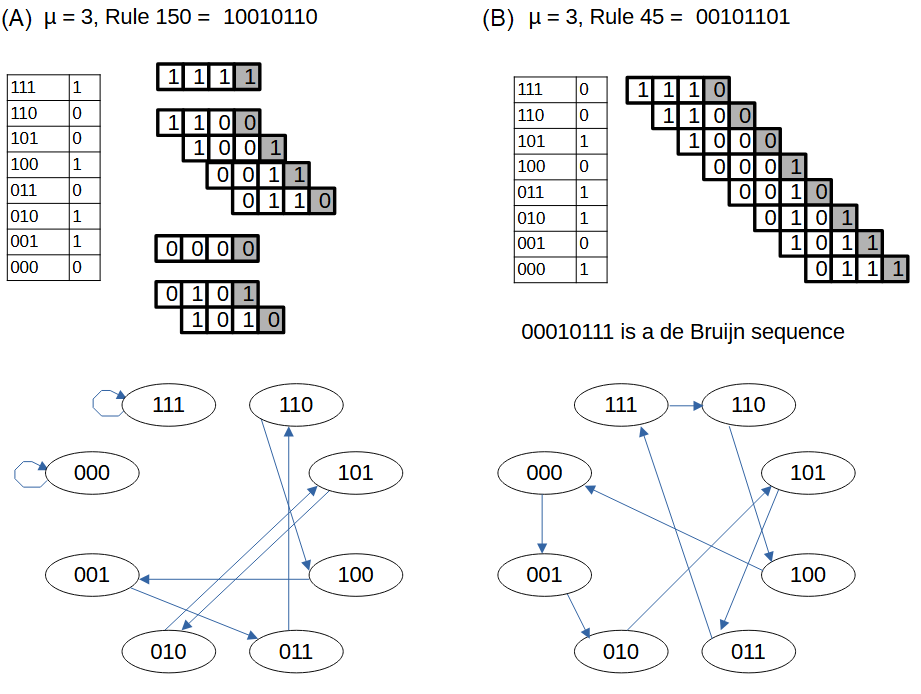}
\caption{Two generating rules with memory $\mu = 3$. 
(\textbf{A}) Rule 150, whose binary representation is $10010110$. This corresponds to the following assignments for each 3-tuple (ordered from $111$ to $000$): $111 \rightarrow 1$, $110 \rightarrow 0$, $101 \rightarrow 0$, $100 \rightarrow 1$, $011 \rightarrow 0$, $010 \rightarrow 1$, $001 \rightarrow 1$, $000 \rightarrow 0$. The application of this rule to four different initial triplets is shown on the right of panel (\textbf{A}). As observed, the resulting sequences are periodic with periods 1, 4, 1, and 2 (from top to bottom). The dynamics of this rule can be visualized as a directed graph, shown at the bottom. Each node represents a possible 3-tuple; a directed edge from node $x$ to node $y$ exists if applying the rule to $x$ yields the last digit of $y$. This graph is not connected and does not contain a Hamiltonian path.
(\textbf{B}) Rule 45, with binary representation $00101101$, corresponds to the truth table shown on the left of panel B. In contrast to Rule 150, applying Rule 45 to any of the $2^3 = 8$ initial triplets yields the same cyclic sequence of maximum period $T_{\text{max}} = 8$, namely, the de Bruijn sequence $s_6 = 00010111$. This rule is thus a de Bruijn rule. The associated graph, built as described above, is a de Bruijn graph because it contains a Hamiltonian path through all $2^3$ nodes.}
\label{fig1}
\end{figure}

To explore periodicity in 0dCA, we adopt combinatorial generating rules analogous to those used in 1dCA \citep{Wolfram1983}, with memory playing the role of spatial interaction. Formally, a generating rule with memory \(\mu\) is a function that maps each of the \(2^\mu\) binary sequences of length \(\mu\) to a single binary output:
\begin{equation}
R(a_1, a_2, \ldots, a_\mu) = a_j
\end{equation}
where \( a_i \in \{0,1\} \) for all \(i=1,2,\ldots,\mu\) (see Figure~\ref{fig1}). Starting from an initial configuration, the rule is applied recursively to generate the digits of the sequence as follows:
\begin{equation}
a_{j+1} = R(a_{j-\mu+1}, a_{j-\mu+2}, \ldots, a_{j})
\end{equation}
which eventually converges to an asymptotic pattern with period \(T\).

\vspace{-4pt}

For example, when \(\mu = 3\), there are \( 2^{2^3} = 256 \) possible rules—corresponding to Wolfram's elementary 1dCA rules. Figure~\ref{fig1}A shows the XOR rule for \( \mu = 3 \), represented by the binary rule string {$10010110$}, which corresponds to rule number 150 in Wolfram’s classification. Applying this rule to the initial configuration \( 0\,1\,0 \) yields
\[
s_5 = 0\, 1\, 0\, 1\, 0\, 1\, 0\, 1\, 0\, \ldots
\]
which has period \( T_5 = 2 \).

In general, for a given \(\mu= 1,2,3, \ldots \), the total number of possible generating rules is \(C(\mu)= 2^{2^\mu} \) ({this number corresponds with the Fermat number minus 1} \cite{A000215}), and the maximum number of binary sequences generated from these rules is $2^{2^\mu + \mu}$. This number grows super-exponentially with \(\mu\), making exhaustive analysis computationally infeasible for large \(\mu\). For instance, when \(\mu = 10\), there are approximately \(10^{308}\) rules, each of which must be applied to 1024 different initial configurations.

This raises the following fundamental question: given a memory value \(\mu\) and an initial configuration, is it possible to predict the asymptotic pattern and period \(T\) of the resulting sequence? While this is tractable for small \(\mu\) via exhaustive enumeration, alternative methods are necessary for large \(\mu\) due to the sheer size of the rule space.


It is worth noting that the 0-dimensional cellular automata (0dCA) with memory, as defined above, are equivalent to sequences generated by a Non-Linear Feedback Shift Register (NLFSR). This type of sequence has been extensively studied, and a vast body of literature exists on the subject \cite{Golomb, Fredricksen, Gao, Zhao}. In particular, significant attention has been devoted to the characterization and computation of shift register sequences with maximum period—commonly known as de Bruijn sequences \cite{deBruijn, Ralston, Huang, Miros}. As we will show in the next section, there exists a one-to-one correspondence between de Bruijn sequences and their generating rules, which we will refer to as de Bruijn rules.

\section{Rules That Generate Maximum Period Sequences: De Bruijn Rules}

\begin{figure}[H]
\includegraphics[width=0.95\textwidth]{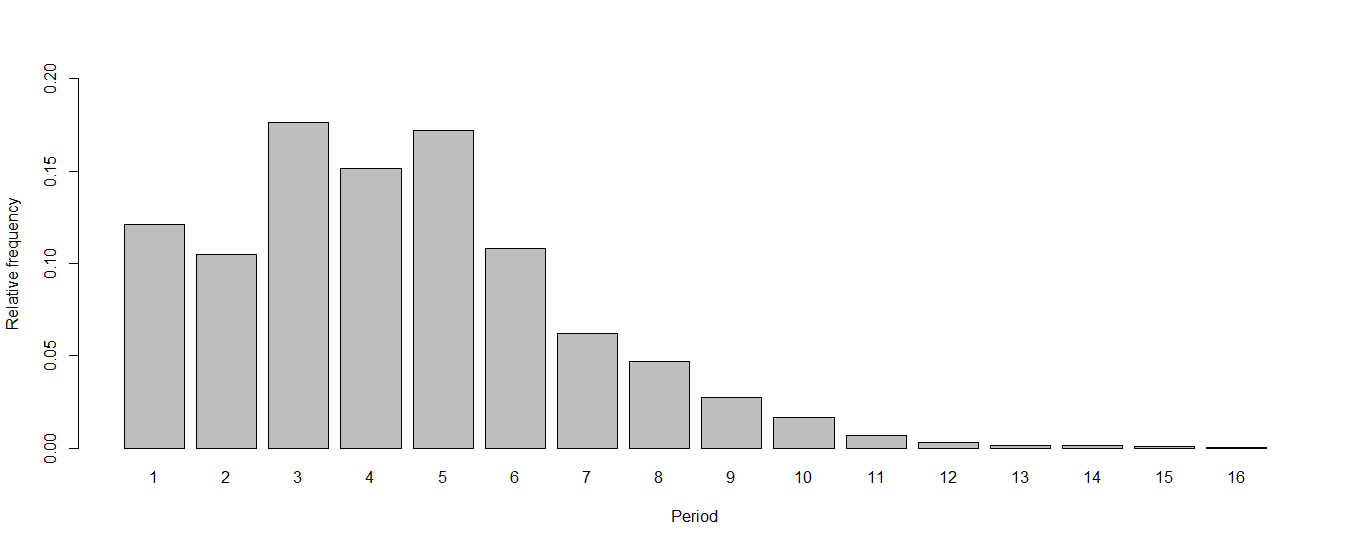}
\caption{Bar chart of the proportion of rules as a function of the maximum period of the sequences that each of them generates over the initial sequences for a memory $\mu = 4$. Note that $T=16$ has the lower occurrence; as a matter of fact, there are $2^{2^3-4}$ de Bruijn rules, which result in a relative frequency of $16/65,536 \approx 0.0002$.}
\label{fig2}
\end{figure}

For a given memory length $\mu$, the maximum period that a generated binary sequence can attain is $T_{\text{max}} = 2^{\mu}$. Naturally, the minimum period is 1, corresponding to sequences that converge to fixed points (e.g., either 0 or 1). For small values of $\mu$, it is feasible to generate all possible sequences and compute their corresponding periods. For $\mu=4$, Table~\ref{tab1} shows the number of rules that generate sequences with periods in between 1 and $2^4$ for each of the $2^4$ initial sequence of length $4$. Note that the distributions of periods are symmetric to middle rows and that the maximum period occurs for only 16 rules, independently of the initial sequence. The proportion of rules as a function of the maximum period of the sequences that each of the $2^{2^4}$ generates over all the initial sequences is depicted in Figure~\ref{fig2} (also for the case $\mu = 4$). It is worthy to remark the nonmonotonous shape of this distribution that exhibits three relative maxima at periods $T=1$, $T=3$, and $T=5$. For $T> 5$, it seems to decrease exponentially.

Sequences with a maximum period, known as de Bruijn sequences, are of particular interest. These are cyclic sequences of length $2^{\mu}$ in which every possible binary substring of length $\mu$ appears exactly once. For small values of $\mu$, it is straightforward to generate all possible de Bruijn sequences. There exists a unique de Bruijn sequence for $\mu = 2$, namely, {{$0011$}}. 
 For $\mu = 3$, there are two such sequences: {$00010111$} and {$00011101$} that are generated by rules 45 and 75, respectively (rule 45 is graphically described in Figure~\ref{fig1}B).

Since these sequences are cyclic, their representations are not unique; by convention, we select the lexicographically least sequence (i.e., the smallest in binary order and decimal number) as the representative of the equivalence class. Specially significative is the lexicographically least sequence for each memory $\mu$ that is referred to as the {{granddaddy}} \mbox{in \cite{Knuth}.}

\begin{table}[H]
\caption{Number of rules that generates sequences with each of the possible periods (columns) that can appear, from 1 to 16, for $\mu = 4$, applied to the $2^4$ initial sequences (rows). Notice that the total number of rules for this $\mu$-value is $2^{2^4}$ = 65,536, which equal the sum of each row.  As it can be observed, the distributions of periods are symmetric to middle rows. Note that the maximum period occurs at the last column ($T=16$) for 16 rules and for every initial condition.}
\label{tab1}
\scriptsize
{\setlength{\tabcolsep}{4pt}
\begin{tabularx}{\textwidth}{rrrrrrrrrrrrrrrrr}
\toprule
\textbf{Initial Sequence} &\textbf{1} & \textbf{2} & \textbf{3} & \textbf{4} & \textbf{5} & \textbf{6} & \textbf{7} & \textbf{8} & \textbf{9} & \textbf{10} & \textbf{11} & \textbf{12} & \textbf{13} & \textbf{14} & \textbf{15} & \textbf{16}\\
\midrule
{{0000}} 
 & {36,096} 
 & 3304 & 4944 & 4734 & 5836 & 3758 & 2472 & 1846 & 1120 &  758 &  332 &  144 &   80  &  48 &   48 &   16 \\ 
{{0001}} & 12,912 & 6608 & 9888 & 9468 & 9624 & 6492 & 3920 & 2924 & 1728 & 1068 &  440 &  208 &   96  &  80  &  64 &   16\\ 
{{0010}} & 11,800 & 8976 & 11,780 & 8457 & 8976 & 5697 & 3552 & 2803 & 1634 &  995  & 412  & 200 &   94  &  80  &  64 &   16 \\ 
{{0011}} & 15,368 & 5320 & 9490 & 9252 & 9296 & 6397 & 3702 & 2981 & 1756 & 1067 &  444 &  207  &  96 &   80  &  64  &  16 \\ 

{{0100}} & 14,568 & 5524 & 11,588 & 8860 & 9190 & 5851 & 3570 & 2834 & 1672 & 1011 &  414 &  200  &  94  &  80  &  64 &   16\\ 
{{0101}} & 10,520 & 16,384 & 8458 & 6494 & 8542 & 5755 & 3620 & 2572 & 1438 &  919 &  406 &  192  &  80 &   76  &  64  &  16 \\ 
{{0110}} & 12,032 & 6428 & 12,398 & 9128 & 9438 & 6043 & 3486 & 2975 & 1714 & 1019  & 422 &  199   & 94  &  80  &  64  &  16 \\ 
{{0111}} & 21,376 & 4956 & 7416 & 8125 & 8242 & 5637 & 3580 & 2641 & 1648 & 1025 &  426 &  208  &  96  &  80 &   64  &  16 \\ 

{{1000}} & 21,376 & 4956 & 7416 & 8125 & 8242 & 5637 & 3580 & 2641 & 1648 & 1025  & 426  & 208   & 96 &   80 &   64   & 16 \\ 
{{1001}} & 12,032 & 6428 & 12,398 & 9128 & 9438 & 6043 & 3486 & 2975 & 1714 & 1019 &  422 &  199  &  94   & 80 &   64  &  16\\ 
{{1010}} & 10,520 & 16,384 & 8458 & 6494 & 8542 & 5755 & 3620 & 2572 & 1438 &  919  & 406 &  192 &   80  &  76 &   64 &   16 \\ 
{{1011}} & 14,568 & 5524 & 11,588 & 8860 & 9190 & 5851 & 3570 & 2834 & 1672 & 1011 &  414 &  200  &  94 &   80   & 64 &   16\\ 

{{1100}} & 15,368 & 5320 &  9490 & 9252 & 9296 & 6397 & 3702 & 2981 & 1756 & 1067  & 444 &  207 &   96 &   80 &  64   & 16\\ 
{{1101}} & 11,800 & 8976 & 11,780 & 8457 & 8976 & 5697 &  3552 & 2803 & 1634 &  995 &  412 &  200  &  94  & 80   & 64 &   16 \\ 
{{1110}} & 12,912 & 6608 & 9888 & 9468 & 9624 & 6492 & 3920 & 2924 & 1728 & 1068 &  440 &  208  &  96 &   80  &  64 &   16\\ 
{{1111}} & 36,096 & 3304 & 4944 & 4734 & 5836 & 3758 & 2472 & 1846 &  1120 &  758  & 332 & 144  &  80 &   48 &   48  &  16\\ 
\bottomrule
\end{tabularx}}
\end{table}

\vspace{-20pt}

\textls[-25]{The number of binary de Bruijn sequences for a given $\mu$ is given by $C(\mu) = 2^{2^{\mu - 1} - \mu}$~\citep{deBruijn, A016031}.} This number grows also superexponentially with $\mu$ and reaches extremely large values even for moderate $\mu$. For example, for $\mu = 10$, $C(10) \approx 10^{151}$. Generating all de Bruijn sequences is a challenging combinatorial problem that remains unsolved in full generality, though various efficient construction algorithms have been proposed (see, for instance, \cite{Hall, deBruijnproject}).

De Bruijn sequences can be generated by applying certain specific rules, independently of the initial configuration of length $\mu$. In other words, there exists a bijection between de Bruijn sequences and their corresponding rules, which we refer to as {{de Bruijn rules}}. 
 Table~\ref{tabX} lists the de Bruijn rules and their de Bruijn sequences for $\mu = 1, 2, 3$, and $4$, expressed both as binary maps and in their corresponding decimal representations.

For $\mu=4$, there are $C(4) = 2^{2^{3} - 4} = 16$ de Bruijn rules and their associated de Bruijn sequences. As it can be seen, the granddaddy sequence is $0000100110101111$. The corresponding granddaddy rule is $3825$, with binary representation $0000111011110001$. For $\mu=5$, the number of de Bruijn sequences is $C(5) = 2^{2^{4} - 5} = 2048$. The granddaddy rule in this case is $00001100111111101111001100000001$ (whose decimal representation is 218,034,945), which generates the granddaddy sequence $00000100011001011011101010011111$. 

\begin{table}[H]
\footnotesize
\caption{Correspondence betweeen de Bruijn rules and de Bruijn sequences for $\mu=1,2,3$ and $4$. The table also shows the Evil Odd Number that divides each de Bruijn rule, leaving the remainder $\phi(\mu)$. The Evil Odd Numbers that correspond to de Bruijn rules for $\mu=5$ are presented in Appendix \ref{App2}.}
\label{tabX}
     
    \begin{tabularx}{\textwidth}{llllll}
    \toprule
        \boldmath{$\mu$}  & \textbf{Evil Odd Number} &\boldmath{$\phi(\mu)$} & \textbf{Rule in Decimal} & \textbf{Rule in Binary}  & \textbf{de Bruijn Sequence} \\ 
		\midrule
        1  & - &-& 1 & 01  & 01 \\ 
        2  & - & 3 & 3 & 0011 & 0011 \\ 
        3  & 3 &15& 45 & 00101101  & 00010111 \\ 
        3  & 5 &15& 75 & 01001011  & 00011101 \\ 
        4  & 3 &255& 765 & 0000001011111101  & 0000101101001111 \\ 
        4  & 9 &255& 2295 & 0000100011110111 & 0000110100101111 \\ 
        4  & 15 &255& 3825 & 0000111011110001  & 0000100110101111 \\ 
        4  & 17 &255& 4335 & 0001000011101111  & 0000111100101101 \\ 
        4  & 27 &255& 6885 & 0001101011100101  & 0000101111001101 \\ 
        4  & 29 &255& 7395 & 0001110011100011  & 0000110101111001 \\ 
        4  & 43 &255& 10,965 & 0010101011010101  & 0000101001101111 \\ 
        4  & 57 &255& 14,535 & 0011100011000111  & 0000110111100101 \\ 
        4  & 65 &255& 16,575 & 0100000010111111  & 0000111101001011 \\ 
        4  & 71 &255& 18,105 & 0100011010111001  & 0000100111101011 \\ 
        4  & 75 &255& 19,125 & 0100101010110101  & 0000101111010011 \\ 
        4  & 83 &255& 21,165 & 0101001010101101  & 0000101100111101 \\ 
        4  & 85 &255& 21,675 & 0101010010101011  & 0000111101011001 \\ 
        4  & 89 &255& 22,695 & 0101100010100111  & 0000110010111101 \\ 
        4  & 99 &255& 25,245 & 0110001010011101  & 0000101001111011 \\ 
        4  & 113 &255& 28,815 & 0111000010001111  & 0000111101100101 \\ 
\bottomrule
    \end{tabularx}
\end{table}

\section{Characterization of de Bruijn Rules}\label{dB}

As stated in the previous section, certain generating rules produce sequences with maximum period—we refer to them as {de Bruijn rules}. Therefore, a complete characterization of such rules emerges as a fundamental goal. The following properties help identify and constrain the set of de Bruijn rules and can be used to systematically search for them:

\begin{itemize}

\item {{Boundary Conditions:}} The binary representations of de Bruijn rules must start with $0$ and end with $1$. This condition arises from the convention in rule ordering, which arranges input strings from the highest binary value ($11\ldots11$) to the lowest ($00\ldots00$) and from the necessity to avoid fixed points (e.g., if $R(11\ldots11) = 1$, the sequence remains constant). This constraint immediately reduces the number of candidate rules by a factor of 4. For example, for $\mu = 4$, the number of potentially valid rules is reduced from $2^{2^4}$ = 65,536 to 16,384.

\item {{Symmetry and Parity:}} The binary representations of de Bruijn rules are symmetric with respect to their midpoint, such that each half is the complement of the other. This symmetry ensures parity: the number of $0$s equals the number of $1$s, although the balance may be broken within each half. Under this constraint, valid de Bruijn rules correspond to multiples of certain Numbers derived from sequences of Evil Odd numbers \citep{A129771}, multiplied by factors denoted as $\phi(\mu)$ (see Appendix \ref{App1}). Let $R_{\mu}$ be a rule for a given $\mu$. Then, 
\begin{equation}\label{symR}
R_{\mu} = \phi(\mu) \, (A+1)
\end{equation}

In Appendix \ref{App1}, we show that the factor $\phi(\mu)$ is
\begin{equation}
\phi(\mu) = 2^{2^{\mu -1}} - 1 
\end{equation}
For instance, $\phi(2) = 3$, $\phi(3) = 15$, $\phi(4) = 255$, and $\phi(5)$ = 65,535.
The consideration of this property causes a huge reduction in the feasible set of the de Bruijn rules, from \mbox{16,384 to} 64 for the case $\mu = 4$.

\item {{Evil Odd Number divisibility:}} It is also empirical evidence that factor $A+1$ in Equation (\ref{symR}) must be an Evil Odd Number for $R_{\mu}$ to be a de Bruijn rule. For each $\mu$, there is a unique $\phi$ that divides $R_{\mu}$, which results in multiple remainders that are Evil Odd Numbers. To consider that, it is convenient to denote this factor as $A(\mu)_k$, representing the Evil Odd Numbers for $k=1,2,\ldots, C(\mu)$. For instance,
\[
A(2,1) = 1 \times 3 = 3, \quad A(3,1) = 3 \times 15 = 45, \quad A(3,2) = 5 \times 15 = 75
\]

Table~\ref{tabX} presents all de Bruijn rules for $\mu = 4$ whose decimal representations are products of an Evil Odd Number and the corresponding $\phi(\mu)$. Note, however, that not all Evil Odd Numbers yield valid de Bruijn rules, and the problem of determining which ones do remains open for larger values of $\mu$. For $\mu = 4$, applying these conditions reduce the feasible set of de Bruijn rules to 32.

\item {{Constrained Position Pairs:}} There exist pairs of positions in the binary representation of de Bruijn rules that cannot simultaneously take the same value (1 for even $\mu$-values and 0 for odd ones). This condition is applied only to the first half of the binary string for symmetry (according to the previous item) and depends on $\mu$ according to a recursive pattern.

Let $p_{\mu}^j$ be the binary value of the $j$ position in a de Bruijn rule of memory $\mu$. For $\mu = 2$, the constrained positions are $p^1_2 = 1$ and $p^2_2 = 2\,p^1_2$, and 
\begin{itemize}[leftmargin=2.5em,labelsep=5.8mm]
    \item If $\mu = 2k + 1$, then $p^1_{2k+1} = p^2_{2k}$ and $p^2_{2k+1} = 2 \, p^1_{2k+1}-1$.
    \item If $\mu = 2\, (k+1)$, then $p^1_{2\,(k+1)} = p^2_{2\, k+1}$ and $p^2_{2 \, (k+1)} = 2\,p^1_{2\,(k+1)}$.
\end{itemize}
for $k = 1, 2, \ldots$. 

This structural condition eliminates $1/4$ of the remaining candidates. For example, for $\mu = 3$, these positions are $p^1_3=2$ and $p^2_3 = 3$. For $\mu=4$, $p^1_4=3$ and $p^2_4 = 6$, and for $\mu=5$, $p^1_5 = 6$ and $p^2_5 = 11$.

Remarkably, for $\mu = 4$, the final count of feasible de Bruijn rules after applying all constraints is 24. In general, for all values of $\mu > 3$, the feasible set exceeds the actual number of de Bruijn rules (see Table~\ref{tab3}).

\item {{Symmetric Rule Invariance:}} If a rule of the form
\[
0\,a_1\,a_2\,\ldots\,a_{k-1}\,a_k\,0\,1\,1-a_1\,1-a_2\,\ldots\,1-a_{k-1}\,1-a_k\,1
\]
is a de Bruijn rule, then its \textit{mirrored} version
\[
0 \,a_k\,a_{k-1}\,\ldots\,a_2\,a_1\,0\,1\,1-a_k\,1-a_{k-1}\,\ldots\,1-a_2\,1-a_1\,1
\]
is also a de Bruijn rule.

This property reflects the inherent symmetry and reversibility in de Bruijn rule structure. It ensures that for each valid de Bruijn rule constructed in this way, a corresponding reverse-complement rule also exists within the de Bruijn set.

\end{itemize}

Let us summarize the reduction procedure that results from the application of these properties for the case $\mu = 4$. There are 65,536 rules that are filtered as follows:
\begin{itemize}[leftmargin=*,labelsep=5.8mm]
\item After applying the boundary conditions, 16,384 rules remain feasible.
\item Applying symmetry, i.e., dividing by $\phi(4) = 255$, 64 rules remain. 
\item The requirement that the remainder is an Evil Odd Number reduces the previous number to 32 rules.
\item Verifying the constrained position pairs reduces the feasible set to 24 rules.
\end{itemize}

From these 24 feasible rules, only 16 are de Bruijn (see Table \ref{tabX}).

The application of the properties stated in the previous paragraphs to the entire set of generating rules significantly reduces the number of feasible rules that can yield de Bruijn sequences (an example can be found in Appendix \ref{App3}). Table~\ref{tab3} presents the total number of rules, the number of de Bruijn sequences, the number of feasible rules after applying the constraints, and the corresponding ratios.  As can be observed, the number of feasible rules that need to be checked to find de Bruijn rules is drastically smaller than the total number of rules for each $\mu$ value. For instance, when $\mu=6$, the reduction factor is approximately $10^{-11}$. In the same case, the ratio between the number of rules that actually yield de Bruijn sequences and the number of feasible rules is around $0.17$. For such low values of $\mu$, a brute-force approach may still be used to generate all de Bruijn sequences. However, a more efficient and structured methodology will be presented in the next section.

\begin{table}[H] 
\scriptsize
\caption{Table resulting from the application of the properties described in Section~\ref{dB}. The second column, $C(\mu)$, indicates the total number of rules for each value of $\mu$. The third column shows the number of feasible rules remaining after applying the constraints detailed in Section~\ref{dB}. The fourth column lists the number of de Bruijn rules. The remaining columns present ratios between these subsets. Particularly noteworthy is the fifth column, which displays the ratio between the feasible subset and the total number of rules. As $\mu$ increases, the reduction in the search space for de Bruijn rules becomes dramatic — for example, for $\mu = 9$, the feasible subset constitutes only about $10^{-79}$ of the full rule set.}
\label{tab3}
 
\begin{tabularx}{\textwidth}{lllllll}
    \toprule
\boldmath{$\mu$} & \boldmath{$C(\mu)$} & \boldmath{$\#$ \textbf{Feasible}} & \boldmath{$\#$ \textbf{de Bruijn}} & \textbf{Feasible/Total} & \textbf{de Bruijn/Total} & \textbf{de Bruijn/Feasible} \\ 
\midrule
2 & 16                    &                & 1                   &                      & 0.0625                &                          \\
3 & 256                   & 2              & 2                   & 0.0078125            & 0.0078125             & 1                        \\ 
4 & 65,536                 & 24             & 16                  & 0.000366211          & 0.00024414           & 0.66666667              \\ 
5 & 4,294,967,296            & 6144           & 2048                & {1.4305} 
 $\times$ 10\textsuperscript{$-6$}        & 4.7683 $\times$ 10\textsuperscript{$-7$}           & 0.33333333              \\ 
6 & 1.84467 $\times$ 10\textsuperscript{$19$}           & 402,653,184      & 67,108,864            & 2.1827 $\times$ 10\textsuperscript{$-11$}          & 3.6379 $\times$ 10\textsuperscript{$-12$}          & 0.16666667              \\ 
7 & 3.40282 $\times$ 10\textsuperscript{$38$}           & 1.7293 $\times$ 10\textsuperscript{$18$}    & 1.4411 $\times$ 10\textsuperscript{$17$}       & 5.0820 $\times$ 10\textsuperscript{$-21$}           & 4.2351 $\times$ 10\textsuperscript{$-22$}         & 0.08333333              \\ 
8 & 1.15792 $\times$ 10\textsuperscript{$77$}          & 3.1901 $\times$ 10\textsuperscript{$37$}    & 1.3292 $\times$ 10\textsuperscript{$36$}        & 2.7550 $\times$ 10\textsuperscript{$-40$}          & 1.1479 $\times$ 10\textsuperscript{$-41$}          & 0.04166667              \\ 
9 & 1.3408 $\times$ 10\textsuperscript{$154$}          & 1.0855 $\times$ 10\textsuperscript{$76$}    & 2.2615 $\times$ 10\textsuperscript{$74$}        & 8.0964 $\times$ 10\textsuperscript{$-79$}          & 1.6867 $\times$ 10\textsuperscript{$-80$}           & 0.02083333             \\
\bottomrule
\end{tabularx}
\end{table}

\section{Neural Networks to Classify de Bruijn Rules}\label{ml}

An alternative approach to identifying de Bruijn rules is to apply machine learning methods. In particular, neural networks are especially well suited for classification tasks~\cite{Bishop}. In this section, we present a neural network model for classifying the feasible rules into two categories: de Bruijn rules (coded by 1) and the rest (0). Although classification based on the sequence period is also possible, it would require accounting for dependence on initial conditions. The classification is performed for $\mu = 5$ and $\mu = 6$ (see Table~\ref{metrics}).

We implemented a binary classification model using a feedforward neural network in \texttt{R} (version 4.5.0)~\cite{Rcore}. The analysis followed these main steps:
\begin{enumerate}
    \item {{Data loading and preprocessing:}}
 The input data consisted of a character string representing a binary sequence and a binary integer label.
    
    \item {{Feature extraction:}} Each rule was split into individual bits transforming the strings into a matrix where each column corresponds to a bit (bit$_1$ to bit$_{2^{\mu}}$). According to the necessary properties of de Bruijn rules (see Section~\ref{dB}), only the first $\mu/2$ bits were retained for further analysis. The first and the $2^{\mu-1}$ bits were also removed because they are necessarily 0.

    \item {{Dataset splitting:}} The data were randomly split into training (80\%) and testing (20\%) subsets to evaluate model performance on unseen data. The validation set was chosen to be $20\%$ of the training set in all cases.

    \item {{Model specification and training:}} A feedforward neural network was constructed {using} 
 the \texttt{keras} package (version 2.15.0) with a TensorFlow (version 2.16.0) backend~\cite{Allaire, Rcore}.
\end{enumerate}

Feedforward neural networks are highly suitable for structured, tabular data, such as the bit vectors extracted from rule representations. They can capture complex, non-linear relationships between input features without requiring explicit feature engineering. In our case, the binary inputs represent discrete features, and their interactions are not trivially captured by simpler linear models. The multi-layer architecture enables hierarchical feature learning, significantly improving classification accuracy.

For the case $\mu=5$, the neural network architecture includes the following:
\begin{itemize}
    \item An input layer with 14 features (bits);
    \item A hidden dense layer with 32 units and ReLU activation;
    \item A second hidden dense layer with 16 units and ReLU activation;
    \item An output layer with 1 unit and sigmoid activation for binary classification.
\end{itemize}

The model was compiled using the Adam optimizer (learning rate $= 0.001$), binary cross-entropy loss function, and accuracy as a performance metric. The dataset includes the complete set of 6144 feasible rules, one-third of which are de Bruijn rules (see Table~\ref{tab3}). Training was performed over 100 epochs with a batch size of 4. Model evaluation was conducted on the test set by calculating accuracy, sensitivity, and specificity (see Table~\ref{metrics}). Class labels were assigned using a threshold of 0.5 on the predicted probabilities. The resulting classifier achieved outstanding performance, with an accuracy exceeding 99\% for $\mu = 5$.

The more challenging case of $\mu = 6$ involves a rule space of size on the order of $10^{19}$, with approximately 67 million de Bruijn rules (see Table~\ref{tab3}). As a matter of fact, we can only use a sample of the total rule space that is randomly analyzed and classified into the two classes. To get this sample of the feasible rule space, we used brute force to generate \mbox{1 million} de Bruijn rules. In parallel, we generated a larger number of non-de Bruijn feasible rules and randomly selected 1 million of them. Note that the ratio of de Bruijn to feasible rules is about 0.17 (Table \ref{tab3}), which means that approximately 6 million feasible rules must be generated by brute force to obtain 1 million de Bruijn rules.

Optimal results were achieved using a slightly deeper neural network with the following architecture:
\begin{itemize}
    \item Three hidden dense layers with 64, 64, and 8 units, respectively;
    \item ReLU activations in all hidden layers;
    \item A sigmoid output unit for binary classification.
\end{itemize}

Training used a batch size of 64 and a learning rate of $0.001$. The dataset consists of a sample of $2 \times 10^6$ feasible rules, half of which are de Bruijn rules. As before, 80\% of the dataset was used for training, with 20\% used for validation and the remainder for testing. A threshold of 0.5 was again applied to the output probabilities to assign class labels. As shown in Table~\ref{metrics}, this model also demonstrated excellent classification metrics.

\begin{table}[H]
\caption{Evaluation metrics for the neural network models applied to classify the de Bruijn rules for memories with $\mu=5$ and $\mu=6$. Total refers to the {sum} 
  TP+FP+TN+FN and corresponds to one fifth of the whole dataset.  The positive class considered {{1}} 
 $\to$ de Bruijn.}
\label{metrics}
    \centering
\begin{tabularx}{\textwidth}{lrr}
    \toprule
\textbf{Metric, Definition} & \boldmath{$\mu = 5$} & \boldmath{$\mu = 6$} \\
\midrule
True Positives (TP) & 397 & 198,563 \\
False Positives (FP) & 3 & 19,839 \\
True Negatives (TN) & 820 & 180,668 \\
False Negatives (FN) & 9 & 930 \\
Accuracy, \small{(TP + TN)/Total} & 0.9902 & 0.9481 \\
Sensitivity (Recall), \small{TP/(TP + FN)} & 0.9778 & 0.9953 \\
Specificity, \small{TN/(TN + FP)} & 0.9964 & 0.9011 \\
Precision (PPV), \small{TP/(TP + FP)} & 0.9925 & 0.9092 \\
Negative Predictive Value (NPV), \small{TN/(TN + FN)} & 0.9891 & 0.9949 \\
Balanced Accuracy, \small{(Sens. + Spec.)/2} & 0.9871 & 0.9482 \\
Detection Rate, \small{TP/Total} & 0.3230 & 0.4964 \\
Detection Prevalence, \small{(TP + FP) / Total} & 0.3255 & 0.5460 \\
True Prevalence, \small{(TP + FN) / Total} & 0.3303 & 0.4987 \\
\bottomrule
\end{tabularx}

\end{table}

Once the neural network model is available, any given rule can be evaluated to determine, with high probability, whether it is a de Bruijn rule. If the prediction is positive, the rule is then applied to generate the corresponding binary sequence, which must ultimately be verified to confirm that it satisfies the de Bruijn sequence properties.

\section{Discussion}

The successive application of updating rules to an initial configuration of $\{0,1\}$ generates a binary sequence that becomes asymptotically periodic. The size of the initial configuration depends on the memory of the rule, denoted by $\mu$, which is the number of digits required to update the next one. For large values of $\mu$, the number of possible generating rules becomes so large that an exhaustive analysis of the resulting patterns is infeasible. Particularly relevant is a specific and highly constrained subset of rules that we named as de Bruijn rules and that generate sequences of maximum period, known as de Bruijn sequences. Although these sequences represent only a tiny fraction of the entire set of possible sequences, their sheer number for large $\mu$ remains enormous, and the complete and effective generation of all such sequences is still an open problem.

In this paper, we have presented a novel approach to compute de Bruijn sequences that combines two complementary methodologies. First, by exploiting structural properties of de Bruijn rules—i.e., those rules that generate maximum period sequences—a drastic reduction in the full set of candidate rules can be achieved. Table~\ref{tab3} summarizes the reduction ratios for several values of $\mu$. Second, we applied a machine learning approach to this feasible subset in order to accurately identify the de Bruijn rules. As shown in Section~\ref{ml}, the use of a classical neural network model for $\mu = 5$ and $6$ allows for a nearly complete classification of the de Bruijn rules in the smaller cases and achieves over 99\% and 94\% accuracy for $\mu = 5$ and $\mu=6$, respectively. Once the de Bruijn rules are identified, the corresponding de Bruijn sequences are straightforward to construct. This would allow one to find the {granddaddy} sequence that represents the lexicographically smallest member for each $\mu$-value.

The rule-based approach for generating de Bruijn sequences presented in this paper is, to the best of our knowledge, unique in the literature. It employs a hybrid methodology that integrates machine learning techniques within the framework of cellular automata. Specifically, we use rules with memory $\mu$ to generate sequences from an initial configuration of $\mu$ bits. Rules that generate sequences with a maximum period $T= 2^{\mu}$—the so-called de Bruijn sequences—are referred to as the de Bruijn rules. By characterizing these rules based on certain empirical properties, we achieve a significant reduction in the overall rule set. However, even after this reduction, a huge set of feasible rules remains for larger $\mu$-values, which still needs to be classified. To address this, we employ a neural network, which has been shown to perform well for $\mu = 5$ and $\mu = 6$. We believe that further development of this machine learning methodology can yield excellent results for computing de Bruijn sequences with much larger $\mu$-values. It is also worth noting that existing methods based on Nonlinear Feedback Shift Registers have yet to definitively solve the problem. Thus, we consider the results presented in this paper a significant advancement in the study and generation of this important class of sequences.

\appendix
\section[\appendixname~\thesection]{}\label{App1}

In this appendix, we show that a natural number $x_n$ whose binary representation of $2\,n$ bits has the form
\[
x_n \equiv [a_1, a_2, \dots, a_n, 1 - a_1, 1 - a_2, \dots, 1 - a_n]
\]
{for $n \in \mathbb{N}$ and $a_i \in \{0,1\}$ are divisible by}
\[
M_n = 2^n - 1
\]

The decimal representation of $x_n$ can be written as

\[
x_n = \sum_{i=1}^{n} a_i \cdot 2^{2n - i} + \sum_{i=1}^{n} (1 - a_i) \cdot 2^{n - i}
\]
which can be manipulated as follows:
\begin{align*}
x_n &= \sum_{i=1}^{n} \left( a_i \cdot 2^{2n - i} + (1 - a_i) \cdot 2^{n - i} \right) \\
&= \sum_{i=1}^{n} \left( a_i (2^{2n - i} - 2^{n - i}) + 2^{n - i} \right) \\
&= \sum_{i=1}^{n} a_i \cdot 2^{n - i} (2^n - 1) + \sum_{i=1}^{n} 2^{n - i}
\end{align*}

The last sum is geometric and can be summed as
\[
 \sum_{i=1}^{n} 2^{n - i} = 2^n - 1.
\]
and, denotating $A = \sum_{i=1}^{n} a_i \cdot 2^{n - i}$, the decimal expression of $x_n$ can be rewritten:
\[
x_n = (2^n - 1)(A + 1)
\]
Therefore, factor $M_n$ divides $x_n$ for $n \in \mathbb{N}$.

This result can be applied to a de Bruijn rule with period $T=2^{\mu}$. In this case, $n = 2^{\mu - 1}$, and, then, $\phi(\mu) = M_{n} = 2^{2^{\mu - 1}} - 1$. In this way, 
we obtain the factors that appear in Table \ref{tabX}:
\begin{itemize}
    \item $n=4 \Rightarrow  \mu = 3 \quad  \text{and} \quad \phi(3) = 2^4 - 1 = 15 = 3 \cdot 5$.
    \item $n=8 \Rightarrow \mu = 4 \quad \text{and} \quad \phi(4) = 2^8 - 1 = 255 = 3 \cdot 5 \cdot 17$.
    \item $n=16 \Rightarrow \mu = 5  \quad \text{and} \quad \phi(5) = 2^{16} - 1 = 65535 = 3 \cdot 5 \cdot 17 \cdot 257$.
    \item $n=32  \Rightarrow \mu = 6 \quad \text{and} \quad \phi(6) = 2^{32} - 1 = 4294967295 = 3 \cdot 5 \cdot 17 \cdot 257 \cdot 65537$.
\end{itemize}

The other factors $A+1$ are all Evil Odd Numbers \cite{A129771} as we have shown empirically for low values of $\mu$ (see Section \ref{dB}).

Therefore, the decimal representation of any de Bruijn rule for any $\mu$ can be decomposed into two terms, one being $\phi(\mu)$ and the other an Evil Odd Number.

\section[\appendixname~\thesection]{}\label{App2}

The decimal representation of the de Bruijn rules for each $\mu$ is divisible for $\phi(\mu)$. The remainder of this division are Evil Odd Numbers. In Table \ref{tabX}, the Evil Odd Numbers corresponding to $\mu = 2,$  $3$,  and $4$ are depicted. In this appendix, we show the Evil Odd Numbers that divide the de Bruijn rules for $\mu =5$. In total, there are 2048 Evil Odd Numbers, one for each de Bruijn rule. 

\begin{table}[H]
\caption{This table continues Table \ref{tabX} and depicts the subset of Evil Odd Numbers for $\mu=5$. The corresponding $\phi$-value is $\phi(5) = 65,535$.}
\label{tab2B}
{\tiny
    \raggedright
{\setlength{\tabcolsep}{4pt}
    \begin{tabularx}{\textwidth}{llllllllllllllllll}
    \toprule
         39 & 2599 & 3751 & 6245 & 7565 & 9127 & 10,203 & 12,729 & 13,977 & 16,431 & 18,993 & 20,127 & 22,689 & 23,941 & 25,531 & 26,573 & 29,173 & 30,311 \\
43 & 2605 & 3763 & 6255 & 7571 & 9139 & 10,205 & 12,771 & 13,983 & 16,433 & 18,999 & 20,143 & 22,695 & 23,951 & 25,533 & 26,585 & 29,177 & 30,323 \\
45 & 2611 & 3789 & 6257 & 7573 & 9141 & 10,221 & 12,777 & 13,999 & 16,439 & 19,003 & 20,155 & 22,701 & 23,963 & 25,581 & 26,659 & 29,223 & 30,341 \\
51 & 2617 & 3801 & 6263 & 7589 & 9199 & 10,233 & 12,785 & 14,011 & 16,551 & 19,053 & 20,175 & 22,713 & 23,965 & 25,593 & 26,665 & 29,229 & 30,353 \\
57 & 2623 & 3807 & 6267 & 7593 & 9211 & 10,403 & 12,837 & 14,021 & 16,555 & 19,065 & 20,187 & 22,719 & 23,969 & 25,601 & 26,683 & 29,235 & 30,359 \\
63 & 2671 & 3823 & 6309 & 7599 & 9213 & 10,409 & 12,849 & 14,031 & 16,557 & 19,071 & 20,231 & 22,755 & 23,975 & 25,607 & 26,727 & 29,241 & 30,375 \\
165 & 2683 & 3835 & 6313 & 7601 & 9219 & 10,427 & 12,855 & 14,033 & 16,563 & 19,111 & 20,237 & 22,757 & 23,981 & 25,613 & 26,731 & 29,247 & 30,387 \\
169 & 2725 & 3845 & 6319 & 7611 & 9225 & 10,471 & 12,903 & 14,039 & 16,569 & 19,117 & 20,243 & 22,767 & 23,987 & 25,625 & 26,733 & 29,285 & 30,407 \\
175 & 2735 & 3855 & 6321 & 7613 & 9243 & 10,475 & 12,915 & 14,043 & 16,575 & 19,123 & 20,245 & 22,769 & 23,989 & 25,631 & 26,739 & 29,295 & 30,419 \\
177 & 2737 & 3857 & 6327 & 7619 & 9255 & 10,477 & 12,967 & 14,055 & 16,679 & 19,129 & 20,249 & 22,775 & 23,993 & 25,635 & 26,745 & 29,297 & 30,469 \\
183 & 2743 & 3867 & 6369 & 7625 & 9259 & 10,483 & 12,973 & 14,067 & 16,683 & 19,135 & 20,261 & 22,779 & 24,001 & 25,637 & 26,751 & 29,303 & 30,479 \\
293 & 2747 & 3869 & 6379 & 7633 & 9261 & 10,489 & 12,979 & 14,087 & 16,685 & 19,183 & 20,273 & 22,823 & 24,007 & 25,647 & 26,913 & 29,307 & 30,481 \\
297 & 2797 & 3885 & 6387 & 7649 & 9267 & 10,495 & 12,985 & 14,099 & 16,691 & 19,195 & 20,303 & 22,827 & 24,013 & 25,649 & 26,923 & 29,349 & 30,491 \\
303 & 2809 & 3897 & 6435 & 7655 & 9273 & 10,657 & 12,991 & 14,101 & 16,693 & 19,239 & 20,315 & 22,829 & 24,019 & 25,655 & 26,937 & 29,361 & 30,493 \\
305 & 2815 & 3917 & 6437 & 7661 & 9279 & 10,667 & 13,029 & 14,117 & 16,805 & 19,245 & 20,317 & 22,835 & 24,021 & 25,659 & 26,981 & 29,367 & 30,509 \\
315 & 2853 & 3929 & 6447 & 7667 & 9287 & 10,681 & 13,039 & 14,129 & 16,809 & 19,251 & 20,333 & 22,837 & 24,025 & 25,669 & 26,985 & 29,415 & 30,521 \\
317 & 2863 & 3975 & 6459 & 7669 & 9291 & 10,725 & 13,041 & 14,149 & 16,815 & 19,253 & 20,345 & 22,883 & 24,077 & 25,673 & 26,991 & 29,427 & 30,535 \\
423 & 2865 & 3981 & 6461 & 7673 & 9293 & 10,729 & 13,047 & 14,161 & 16,817 & 19,257 & 20,357 & 22,889 & 24,089 & 25,679 & 26,993 & 29,477 & 30,541 \\
427 & 2875 & 3987 & 6497 & 7685 & 9299 & 10,735 & 13,051 & 14,213 & 16,827 & 19,311 & 20,367 & 22,897 & 24,095 & 25,681 & 27,003 & 29,487 & 30,547 \\
429 & 2877 & 3989 & 6503 & 7697 & 9305 & 10,737 & 13,095 & 14,223 & 16,829 & 19,323 & 20,369 & 22,947 & 24,111 & 25,687 & 27,005 & 29,489 & 30,549 \\
435 & 2925 & 3993 & 6509 & 7703 & 9311 & 10,747 & 13,107 & 14,225 & 16,941 & 19,325 & 20,379 & 22,949 & 24,123 & 25,703 & 27,175 & 29,499 & 30,553 \\
437 & 2937 & 4005 & 6515 & 7719 & 9345 & 10,749 & 13,109 & 14,235 & 16,953 & 19,365 & 20,381 & 22,959 & 24,133 & 25,707 & 27,181 & 29,501 & 30,565 \\
559 & 2983 & 4017 & 6517 & 7731 & 9351 & 10,919 & 13,157 & 14,237 & 16,959 & 19,375 & 20,397 & 22,971 & 24,143 & 25,709 & 27,187 & 29,543 & 30,577 \\
571 & 2989 & 4047 & 6521 & 7751 & 9357 & 10,925 & 13,169 & 14,253 & 17,071 & 19,377 & 20,409 & 22,973 & 24,145 & 25,715 & 27,193 & 29,549 & 30,599 \\
685 & 2995 & 4059 & 6567 & 7763 & 9369 & 10,931 & 13,221 & 14,265 & 17,083 & 19,387 & 20,429 & 23,009 & 24,151 & 25,721 & 27,199 & 29,555 & 30,611 \\
697 & 2997 & 4061 & 6571 & 7821 & 9375 & 10,937 & 13,231 & 14,279 & 17,199 & 19,389 & 20,441 & 23,015 & 24,155 & 25,727 & 27,247 & 29,557 & 30,613 \\
703 & 3001 & 4077 & 6573 & 7833 & 9379 & 10,943 & 13,233 & 14,285 & 17,211 & 19,437 & 20,513 & 23,021 & 24,167 & 25,731 & 27,259 & 29,561 & 30,629 \\
813 & 3055 & 4089 & 6579 & 7839 & 9381 & 10,991 & 13,243 & 14,291 & 17,213 & 19,449 & 20,523 & 23,027 & 24,179 & 25,737 & 27,429 & 29,607 & 30,641 \\
825 & 3067 & 4131 & 6581 & 7855 & 9391 & 11,003 & 13,245 & 14,293 & 17,325 & 19,457 & 20,531 & 23,029 & 24,197 & 25,755 & 27,439 & 29,619 & 30,661 \\
943 & 3069 & 4133 & 6627 & 7867 & 9393 & 11,173 & 13,287 & 14,297 & 17,337 & 19,467 & 20,643 & 23,033 & 24,209 & 25,767 & 27,441 & 29,621 & 30,673 \\
955 & 3075 & 4143 & 6633 & 7877 & 9399 & 11,183 & 13,293 & 14,309 & 17,413 & 19,475 & 20,645 & 23,085 & 24,215 & 25,771 & 27,451 & 29,669 & 30,753 \\
957 & 3081 & 4145 & 6641 & 7887 & 9403 & 11,185 & 13,299 & 14,321 & 17,417 & 19,491 & 20,655 & 23,097 & 24,231 & 25,773 & 27,453 & 29,681 & 30,759 \\
1031 & 3099 & 4151 & 6693 & 7889 & 9413 & 11,195 & 13,301 & 14,497 & 17,423 & 19,493 & 20,657 & 23,103 & 24,243 & 25,779 & 27,501 & 29,699 & 30,765 \\
1035 & 3111 & 4155 & 6705 & 7895 & 9417 & 11,197 & 13,305 & 14,503 & 17,425 & 19,503 & 20,663 & 23,141 & 24,263 & 25,785 & 27,513 & 29,705 & 30,777 \\
1037 & 3115 & 4257 & 6711 & 7899 & 9423 & 11,245 & 13,313 & 14,509 & 17,431 & 19,505 & 20,667 & 23,151 & 24,275 & 25,791 & 27,651 & 29,723 & 30,783 \\
1043 & 3117 & 4267 & 6759 & 7911 & 9425 & 11,257 & 13,319 & 14,521 & 17,447 & 19,511 & 20,771 & 23,153 & 24,335 & 25,799 & 27,657 & 29,735 & 30,819 \\
1049 & 3123 & 4275 & 6771 & 7923 & 9431 & 11,395 & 13,325 & 14,527 & 17,451 & 19,515 & 20,777 & 23,159 & 24,347 & 25,803 & 27,675 & 29,739 & 30,821 \\
1055 & 3129 & 4385 & 6829 & 7943 & 9447 & 11,401 & 13,337 & 14,563 & 17,453 & 19,525 & 20,785 & 23,163 & 24,349 & 25,805 & 27,687 & 29,741 & 30,831 \\
1157 & 3135 & 4391 & 6841 & 7955 & 9451 & 11,419 & 13,343 & 14,565 & 17,459 & 19,529 & 20,897 & 23,205 & 24,365 & 25,811 & 27,691 & 29,747 & 30,833 \\
1161 & 3143 & 4397 & 6847 & 7957 & 9453 & 11,431 & 13,347 & 14,575 & 17,465 & 19,535 & 20,903 & 23,217 & 24,377 & 25,817 & 27,693 & 29,753 & 30,839 \\
1167 & 3147 & 4403 & 6885 & 7973 & 9459 & 11,435 & 13,349 & 14,577 & 17,471 & 19,537 & 20,909 & 23,223 & 24,391 & 25,823 & 27,699 & 29,759 & 30,843 \\
1169 & 3149 & 4405 & 6895 & 7985 & 9465 & 11,437 & 13,359 & 14,583 & 17,543 & 19,543 & 20,915 & 23,271 & 24,397 & 25,859 & 27,705 & 29,761 & 31,011 \\
1175 & 3155 & 4409 & 6897 & 8005 & 9471 & 11,443 & 13,361 & 14,587 & 17,547 & 19,559 & 20,917 & 23,283 & 24,403 & 25,861 & 27,711 & 29,771 & 31,013 \\
1191 & 3161 & 4515 & 6903 & 8017 & 9473 & 11,449 & 13,367 & 14,755 & 17,549 & 19,563 & 20,921 & 23,343 & 24,405 & 25,871 & 27,719 & 29,779 & 31,023 \\
1195 & 3167 & 4521 & 6907 & 8079 & 9483 & 11,455 & 13,371 & 14,757 & 17,555 & 19,565 & 21,029 & 23,355 & 24,409 & 25,883 & 27,723 & 29,795 & 31,035 \\
1197 & 3201 & 4529 & 6951 & 8091 & 9497 & 11,463 & 13,379 & 14,767 & 17,561 & 19,571 & 21,039 & 23,357 & 24,421 & 25,885 & 27,725 & 29,797 & 31,037 \\
1203 & 3211 & 4647 & 6963 & 8093 & 9509 & 11,467 & 13,381 & 14,779 & 17,567 & 19,577 & 21,041 & 23,399 & 24,433 & 25,889 & 27,731 & 29,807 & 31,073 \\
1209 & 3219 & 4659 & 6965 & 8109 & 9513 & 11,469 & 13,391 & 14,781 & 17,671 & 19,583 & 21,047 & 23,405 & 24,455 & 25,895 & 27,737 & 29,809 & 31,079 \\
1215 & 3235 & 4773 & 7013 & 8121 & 9519 & 11,475 & 13,393 & 14,817 & 17,675 & 19,587 & 21,051 & 23,411 & 24,467 & 25,901 & 27,743 & 29,815 & 31,085 \\
1285 & 3237 & 4783 & 7025 & 8135 & 9521 & 11,481 & 13,399 & 14,823 & 17,677 & 19,593 & 21,159 & 23,413 & 24,469 & 25,907 & 27,905 & 29,819 & 31,091 \\
1289 & 3247 & 4785 & 7087 & 8141 & 9531 & 11,487 & 13,403 & 14,829 & 17,683 & 19,611 & 21,171 & 23,417 & 24,485 & 25,909 & 27,915 & 29,825 & 31,093 \\
1295 & 3249 & 4791 & 7099 & 8147 & 9533 & 11,649 & 13,443 & 14,835 & 17,685 & 19,623 & 21,287 & 23,463 & 24,497 & 25,913 & 27,929 & 29,831 & 31,097 \\
1297 & 3255 & 4795 & 7101 & 8149 & 9541 & 11,659 & 13,449 & 14,837 & 17,701 & 19,627 & 21,293 & 23,475 & 24,517 & 25,927 & 27,941 & 29,837 & 31,269 \\
1307 & 3259 & 4901 & 7143 & 8153 & 9545 & 11,673 & 13,467 & 14,841 & 17,705 & 19,629 & 21,299 & 23,477 & 24,529 & 25,931 & 27,945 & 29,849 & 31,281 \\
1309 & 3269 & 4913 & 7149 & 8165 & 9551 & 11,685 & 13,479 & 15,013 & 17,711 & 19,635 & 21,301 & 23,525 & 24,609 & 25,933 & 27,951 & 29,855 & 31,287 \\
1415 & 3273 & 5031 & 7155 & 8177 & 9553 & 11,689 & 13,483 & 15,025 & 17,713 & 19,641 & 21,305 & 23,537 & 24,615 & 25,939 & 27,953 & 29,859 & 31,335 \\
1419 & 3279 & 5037 & 7157 & 8227 & 9563 & 11,695 & 13,485 & 15,031 & 17,723 & 19,647 & 21,413 & 23,557 & 24,621 & 25,941 & 27,963 & 29,861 & 31,347 \\
1421 & 3281 & 5043 & 7161 & 8233 & 9565 & 11,697 & 13,491 & 15,079 & 17,725 & 19,655 & 21,425 & 23,561 & 24,633 & 25,957 & 27,965 & 29,871 & 31,527 \\
1427 & 3287 & 5045 & 7169 & 8251 & 9603 & 11,707 & 13,497 & 15,091 & 17,797 & 19,659 & 21,505 & 23,567 & 24,639 & 25,961 & 27,973 & 29,873 & 31,539 \\
1429 & 3303 & 5049 & 7175 & 8295 & 9605 & 11,709 & 13,503 & 15,271 & 17,801 & 19,661 & 21,515 & 23,569 & 24,677 & 25,967 & 27,977 & 29,879 & 31,541 \\
1445 & 3307 & 5123 & 7181 & 8299 & 9615 & 11,717 & 13,505 & 15,283 & 17,807 & 19,667 & 21,523 & 23,575 & 24,681 & 25,969 & 27,983 & 29,883 & 31,589 \\
1449 & 3309 & 5125 & 7193 & 8301 & 9627 & 11,721 & 13,515 & 15,285 & 17,809 & 19,673 & 21,539 & 23,591 & 24,687 & 25,979 & 27,985 & 29,891 & 31,601 \\
1455 & 3315 & 5135 & 7199 & 8307 & 9629 & 11,727 & 13,523 & 15,333 & 17,819 & 19,679 & 21,541 & 23,595 & 24,689 & 25,981 & 27,995 & 29,893 & 31,745 \\
1457 & 3321 & 5137 & 7203 & 8313 & 9633 & 11,729 & 13,539 & 15,345 & 17,821 & 19,715 & 21,551 & 23,597 & 24,695 & 25,985 & 27,997 & 29,903 & 31,751 \\
1467 & 3327 & 5143 & 7205 & 8319 & 9639 & 11,739 & 13,541 & 15,489 & 17,933 & 19,721 & 21,553 & 23,603 & 24,739 & 25,995 & 28,167 & 29,905 & 31,757 \\
1469 & 3329 & 5147 & 7215 & 8353 & 9645 & 11,741 & 13,551 & 15,495 & 17,945 & 19,729 & 21,559 & 23,609 & 24,745 & 26,009 & 28,173 & 29,911 & 31,769 \\
1551 & 3339 & 5249 & 7217 & 8359 & 9651 & 11,911 & 13,553 & 15,501 & 17,951 & 19,745 & 21,563 & 23,615 & 24,763 & 26,021 & 28,179 & 29,915 & 31,775 \\
1563 & 3353 & 5259 & 7223 & 8365 & 9653 & 11,917 & 13,559 & 15,513 & 17,967 & 19,751 & 21,635 & 23,617 & 24,807 & 26,025 & 28,185 & 29,953 & 31,779 \\
1677 & 3365 & 5267 & 7227 & 8377 & 9657 & 11,923 & 13,563 & 15,519 & 17,979 & 19,757 & 21,637 & 23,627 & 24,811 & 26,031 & 28,191 & 29,963 & 31,781 \\
1689 & 3369 & 5283 & 7235 & 8383 & 9671 & 11,929 & 13,571 & 15,523 & 18,063 & 19,763 & 21,647 & 23,635 & 24,813 & 26,033 & 28,207 & 29,977 & 31,791 \\
1695 & 3375 & 5285 & 7237 & 8421 & 9675 & 11,935 & 13,573 & 15,525 & 18,075 & 19,765 & 21,649 & 23,651 & 24,819 & 26,043 & 28,219 & 29,989 & 31,793 \\
1711 & 3377 & 5295 & 7247 & 8425 & 9677 & 11,951 & 13,583 & 15,535 & 18,191 & 19,769 & 21,655 & 23,653 & 24,825 & 26,045 & 28,239 & 29,993 & 31,799 \\
1723 & 3387 & 5297 & 7249 & 8431 & 9683 & 11,963 & 13,595 & 15,537 & 18,203 & 19,783 & 21,659 & 23,663 & 24,831 & 26,053 & 28,251 & 29,999 & 31,803 \\
1805 & 3389 & 5303 & 7255 & 8433 & 9685 & 11,983 & 13,597 & 15,543 & 18,205 & 19,787 & 21,763 & 23,665 & 24,867 & 26,057 & 28,421 & 30,001 & 31,811 \\
1817 & 3397 & 5307 & 7259 & 8439 & 9701 & 11,995 & 13,601 & 15,547 & 18,221 & 19,789 & 21,769 & 23,671 & 24,869 & 26,063 & 28,431 & 30,011 & 31,813 \\
1935 & 3401 & 5377 & 7301 & 8481 & 9705 & 12,165 & 13,607 & 15,555 & 18,233 & 19,795 & 21,777 & 23,675 & 24,879 & 26,065 & 28,433 & 30,013 & 31,823 \\
1947 & 3407 & 5383 & 7305 & 8491 & 9711 & 12,175 & 13,613 & 15,557 & 18,317 & 19,797 & 21,793 & 23,681 & 24,891 & 26,075 & 28,443 & 30,019 & 31,825 \\
1949 & 3409 & 5389 & 7311 & 8505 & 9713 & 12,177 & 13,619 & 15,567 & 18,329 & 19,813 & 21,799 & 23,687 & 24,893 & 26,077 & 28,445 & 30,025 & 31,831 \\
1965 & 3419 & 5395 & 7313 & 8549 & 9723 & 12,187 & 13,621 & 15,569 & 18,465 & 19,817 & 21,805 & 23,693 & 24,935 & 26,117 & 28,461 & 30,033 & 31,835 \\
1977 & 3421 & 5397 & 7319 & 8553 & 9725 & 12,189 & 13,625 & 15,575 & 18,475 & 19,823 & 21,811 & 23,705 & 24,939 & 26,129 & 28,473 & 30,049 & 32,003 \\
2083 & 3459 & 5401 & 7335 & 8559 & 9735 & 12,205 & 13,633 & 15,579 & 18,483 & 19,825 & 21,813 & 23,711 & 24,941 & 26,135 & 28,493 & 30,055 & 32,005 \\

\bottomrule
    \end{tabularx}
}
}
\end{table}

\begin{table}[H]\ContinuedFloat
\caption{{\em Cont.}}
{\tiny
    \raggedright
{\setlength{\tabcolsep}{4pt}
    \begin{tabularx}{\textwidth}{llllllllllllllllll}
    \toprule

2089 & 3465 & 5507 & 7339 & 8561 & 9741 & 12,217 & 13,639 & 15,747 & 18,533 & 19,835 & 21,817 & 23,715 & 24,947 & 26,151 & 28,505 & 30,061 & 32,015 \\
2107 & 3473 & 5513 & 7341 & 8571 & 9747 & 12,237 & 13,645 & 15,749 & 18,537 & 19,837 & 21,889 & 23,717 & 24,949 & 26,163 & 28,707 & 30,067 & 32,027 \\
2151 & 3489 & 5521 & 7347 & 8573 & 9753 & 12,249 & 13,651 & 15,759 & 18,543 & 19,841 & 21,895 & 23,727 & 24,993 & 26,189 & 28,713 & 30,069 & 32,029 \\
2155 & 3495 & 5537 & 7353 & 8611 & 9759 & 12,321 & 13,653 & 15,771 & 18,545 & 19,851 & 21,901 & 23,729 & 25,003 & 26,201 & 28,731 & 30,073 & 32,033 \\
2157 & 3501 & 5543 & 7359 & 8613 & 9775 & 12,327 & 13,657 & 15,773 & 18,551 & 19,865 & 21,907 & 23,735 & 25,017 & 26,207 & 28,769 & 30,083 & 32,039 \\
2163 & 3507 & 5549 & 7361 & 8623 & 9787 & 12,333 & 13,697 & 15,777 & 18,595 & 19,877 & 21,909 & 23,739 & 25,061 & 26,223 & 28,779 & 30,085 & 32,045 \\
2169 & 3509 & 5555 & 7371 & 8635 & 9807 & 12,345 & 13,707 & 15,783 & 18,601 & 19,881 & 21,913 & 23,747 & 25,065 & 26,235 & 28,787 & 30,095 & 32,051 \\
2175 & 3513 & 5557 & 7379 & 8637 & 9819 & 12,351 & 13,721 & 15,789 & 18,619 & 19,887 & 22,021 & 23,749 & 25,071 & 26,247 & 28,833 & 30,107 & 32,053 \\
2209 & 3527 & 5561 & 7395 & 8679 & 9861 & 12,387 & 13,733 & 15,795 & 18,663 & 19,889 & 22,031 & 23,759 & 25,073 & 26,253 & 28,839 & 30,109 & 32,057 \\
2219 & 3531 & 5639 & 7397 & 8683 & 9873 & 12,389 & 13,737 & 15,797 & 18,667 & 19,899 & 22,033 & 23,761 & 25,083 & 26,259 & 28,845 & 30,113 & 32,065 \\
2227 & 3533 & 5651 & 7407 & 8685 & 9879 & 12,399 & 13,743 & 15,801 & 18,669 & 19,901 & 22,039 & 23,767 & 25,085 & 26,265 & 28,857 & 30,119 & 32,071 \\
2277 & 3539 & 5765 & 7409 & 8691 & 9895 & 12,401 & 13,745 & 15,809 & 18,675 & 19,909 & 22,043 & 23,771 & 25,125 & 26,271 & 28,863 & 30,125 & 32,077 \\
2281 & 3541 & 5775 & 7415 & 8693 & 9907 & 12,407 & 13,755 & 15,815 & 18,681 & 19,913 & 22,055 & 23,815 & 25,137 & 26,287 & 28,899 & 30,131 & 32,083 \\
2287 & 3557 & 5777 & 7419 & 8743 & 9933 & 12,411 & 13,757 & 15,821 & 18,687 & 19,919 & 22,067 & 23,819 & 25,143 & 26,299 & 28,901 & 30,133 & 32,085 \\
2289 & 3561 & 5783 & 7427 & 8749 & 9945 & 12,451 & 13,763 & 15,827 & 18,723 & 19,921 & 22,151 & 23,821 & 25,197 & 26,319 & 28,911 & 30,137 & 32,089 \\
2295 & 3567 & 5787 & 7429 & 8755 & 9951 & 12,457 & 13,769 & 15,829 & 18,729 & 19,931 & 22,163 & 23,827 & 25,209 & 26,331 & 28,913 & 30,145 & 32,261 \\
2337 & 3569 & 5799 & 7439 & 8761 & 9967 & 12,475 & 13,777 & 15,833 & 18,737 & 19,933 & 22,279 & 23,829 & 25,215 & 26,375 & 28,919 & 30,151 & 32,273 \\
2347 & 3579 & 5811 & 7451 & 8767 & 9979 & 12,513 & 13,793 & 16,005 & 18,791 & 19,973 & 22,285 & 23,845 & 25,255 & 26,387 & 28,923 & 30,157 & 32,279 \\
2361 & 3581 & 5893 & 7453 & 8815 & 9989 & 12,523 & 13,799 & 16,017 & 18,795 & 19,983 & 22,291 & 23,849 & 25,261 & 26,389 & 28,961 & 30,163 & 32,295 \\
2405 & 3591 & 5905 & 7457 & 8827 & 9999 & 12,531 & 13,805 & 16,023 & 18,797 & 19,985 & 22,293 & 23,855 & 25,267 & 26,405 & 28,971 & 30,165 & 32,307 \\
2409 & 3597 & 6023 & 7463 & 8869 & 10,001 & 12,579 & 13,811 & 16,039 & 18,803 & 19,991 & 22,297 & 23,857 & 25,273 & 26,417 & 28,985 & 30,169 & 32,327 \\
2415 & 3603 & 6029 & 7469 & 8881 & 10,011 & 12,581 & 13,813 & 16,051 & 18,805 & 19,995 & 22,309 & 23,867 & 25,279 & 26,447 & 29,027 & 30,215 & 32,339 \\
2417 & 3609 & 6035 & 7475 & 8887 & 10,013 & 12,591 & 13,817 & 16,071 & 18,849 & 20,007 & 22,321 & 23,869 & 25,327 & 26,459 & 29,033 & 30,221 & 32,519 \\
2427 & 3615 & 6037 & 7477 & 8941 & 10,029 & 12,603 & 13,829 & 16,083 & 18,859 & 20,019 & 22,405 & 23,875 & 25,339 & 26,461 & 29,041 & 30,227 & 32,531 \\
2429 & 3631 & 6041 & 7481 & 8953 & 10,041 & 12,605 & 13,841 & 16,263 & 18,873 & 20,045 & 22,417 & 23,881 & 25,383 & 26,477 & 29,091 & 30,233 & 32,533 \\
2467 & 3643 & 6053 & 7489 & 8959 & 10,061 & 12,641 & 13,847 & 16,275 & 18,917 & 20,057 & 22,565 & 23,889 & 25,395 & 26,489 & 29,093 & 30,239 & 32,549 \\
2473 & 3663 & 6065 & 7495 & 8997 & 10,073 & 12,647 & 13,863 & 16,277 & 18,921 & 20,063 & 22,569 & 23,905 & 25,397 & 26,501 & 29,103 & 30,255 & 32,561 \\
2481 & 3675 & 6177 & 7501 & 9007 & 10,119 & 12,653 & 13,875 & 16,293 & 18,927 & 20,079 & 22,575 & 23,911 & 25,455 & 26,511 & 29,115 & 30,267 & 32,581 \\
2535 & 3717 & 6183 & 7507 & 9009 & 10,131 & 12,659 & 13,895 & 16,305 & 18,929 & 20,091 & 22,577 & 23,917 & 25,467 & 26,513 & 29,117 & 30,277 & 32,593 \\
2539 & 3727 & 6189 & 7509 & 9019 & 10,133 & 12,661 & 13,907 & 16,325 & 18,939 & 20,103 & 22,583 & 23,923 & 25,469 & 26,523 & 29,153 & 30,287 & ~ \\
2541 & 3729 & 6201 & 7513 & 9021 & 10,149 & 12,665 & 13,959 & 16,337 & 18,941 & 20,109 & 22,625 & 23,925 & 25,509 & 26,525 & 29,159 & 30,289 & ~ \\
2547 & 3735 & 6207 & 7559 & 9069 & 10,161 & 12,705 & 13,965 & 16,421 & 18,981 & 20,115 & 22,635 & 23,929 & 25,519 & 26,541 & 29,165 & 30,295 & ~ \\
2549 & 3739 & 6243 & 7563 & 9081 & 10,191 & 12,715 & 13,971 & 16,425 & 18,991 & 20,121 & 22,643 & 23,939 & 25,521 & 26,553 & 29,171 & 30,299 & ~ \\
\bottomrule
    \end{tabularx}
}
}
\end{table}

\section[\appendixname~\thesection]{}\label{App3}

As an application of the methodology presented in this paper, we can generate sequences with large periods, $T=2^{\mu}$, by tuning the memory parameter $\mu$. We have proven that, once $\mu$ is chosen, all de Bruijn rules share the same divisor: $\phi(\mu) = 2^{2^{\mu-1}}-1$, and the remainders are Evil Odd Numbers. Therefore, to obtain a candidate to de Bruijn rule, take a very large Evil Odd Number and multiply it by $\phi(\mu)$. Check the resulting rule and, if valid, apply the rule to any initial binary sequence of length $\mu$ to obtain the de Bruijn sequence. 

Due to computational limitations, we can only offer an example for $\mu = 8$, which provides a sequence of period of $T=256$. For this case,
\[
\phi(8) = 2^{2^7} - 1 = 340282366920938463463374607431768211455
\]
 and take the Evil Odd Number 155640396308704138405661716133499575. 

Multiply both to obtain the decimal representation of the de Bruijn rule:

\[
R_8 = 52961682444438738040038706846647584952275525588128484521019054507752631625
\]

The binary representation of this de Bruijn rule {is} 
\begin{verbatim}
0000000000011101111110011010100111111011100101001010000100111111
1001100000011011011001001111010110000001000001001101001010110110
1111111111100010000001100101011000000100011010110101111011000000
0110011111100100100110110000101001111110111110110010110101001001
\end{verbatim}

Applying this rule to the initial configuration 00000000 yields the following de Bruijn sequence:
\begin{verbatim}
0000000010001111101100111000111011011100100010000110000001110011
0001011010101011011000110011011010011010000010101110100010100111
1000100111010111100111111010010111000001101111010100100110010100
0110101100001111011101111111100001011111001011001001010100001001
\end{verbatim}

This sequence of period $T=256$ repeats infinitely under the application of the above de Bruijn rule.

\bibliographystyle{unsrt}  


\end{document}